\documentstyle[12pt,epsfig,graphics]{article}
\begin{document}\bibliographystyle{plain}\begin{titlepage}
\renewcommand{\thefootnote}{\fnsymbol{footnote}}\hfill\begin{tabular}{l}
HEPHY-PUB 740/01\\UWThPh-2001-27\\CUQM-86\\hep-th/0110220\\July
2001\end{tabular}\\[.3cm]\Large\begin{center}{\bf DISCRETE SPECTRA OF
SEMIRELATIVISTIC HAMILTONIANS FROM ENVELOPE THEORY}\\\vspace{0.5cm}\large{\bf
Richard L. HALL\footnote[3]{\normalsize\ {\em E-mail address\/}:
rhall@mathstat.concordia.ca}}\\[.1cm]\normalsize Department of Mathematics
and Statistics, Concordia University,\\1455 de Maisonneuve Boulevard West,
Montr\'eal, Qu\'ebec, Canada H3G 1M8\\[0.4cm]\large{\bf Wolfgang
LUCHA\footnote[1]{\normalsize\ {\em E-mail address\/}:
wolfgang.lucha@oeaw.ac.at}}\\[.1cm]\normalsize Institut f\"ur
Hochenergiephysik, \"Osterreichische Akademie der
Wissenschaften,\\Nikolsdorfergasse 18, A-1050 Wien,
Austria\\[0.4cm]\large{\bf Franz F.~SCH\"OBERL\footnote[2]{\normalsize\ {\em
E-mail address\/}: franz.schoeberl@univie.ac.at}}\\[.1cm]\normalsize Institut
f\"ur Theoretische Physik, Universit\"at Wien,\\Boltzmanngasse 5, A-1090
Wien, Austria\vfill {\normalsize\bf Abstract}\end{center}\normalsize We
analyze the (discrete) spectrum of the semirelativistic ``spinless-Salpeter''
Hamiltonian$$H=\beta\sqrt{m^2+{\bf p}^2}+V(r)\ ,\quad\beta>0\ ,$$where $V(r)$
is an attractive, spherically symmetric potential in three dimensions. In
order~to locate the eigenvalues of $H,$ we extend the ``envelope theory,''
originally formulated only~for nonrelativistic Schr\"odinger operators, to
the case of Hamiltonians involving the relativistic kinetic-energy operator.
If $V(r)$ is a convex transformation of the Coulomb potential $-1/r$ and a
concave transformation of the harmonic-oscillator potential $r^2$, both upper
and lower bounds on the discrete eigenvalues of $H$ can be constructed, which
may all be expressed~in the
form$$E=\min_{r>0}\left[\beta\sqrt{m^2+\frac{P^2}{r^2}}+V(r)\right]$$for
suitable values of the numbers $P$ here provided. At the critical point, the
relative growth to the Coulomb potential $h(r)=-1/r$ must be bounded by ${\rm
d}V/{\rm d}h<2\beta/\pi.$\\[3ex]{\em PACS numbers\/}: 03.65.Ge, 03.65.Pm,
11.10.St\renewcommand{\thefootnote}{\arabic{footnote}}\end{titlepage}

\normalsize

\section{Introduction}We study the semirelativistic (so-called
``spinless-Salpeter'') Hamiltonian\begin{equation}H=\beta\sqrt{m^2+{\bf
p}^2}+V(r)\ ,\quad\beta>0\ ,\label{Eq:SH}\end{equation}in which $V(r)$ is a
central potential in three dimensions. The eigenvalue equation~of this
operator is called the ``spinless Salpeter equation.'' This equation of
motion arises as~a well-defined standard approximation to the Bethe--Salpeter
formalism \cite{Salpeter51} for the description of bound states within a
(relativistic) quantum field theory and is arrived at by the following
simplifying steps:\begin{enumerate}\item Eliminate all timelike variables by
assuming the Bethe--Salpeter kernel that describes the interactions between
the bound-state constituents to be static, i.e., instantaneous; the result of
this reduction step is called the ``instantaneous Bethe--Salpeter equation''
or the ``Salpeter equation'' \cite{Salpeter52}.\item Neglect the spin of the
bound-state constituents, assume the Bethe--Salpeter kernel~to be of
convolution type (as is frequently the case), and consider merely
positive-energy solutions $\psi,$ in order to arrive at the so-called
``spinless Salpeter equation'' $H\psi=E\psi,$ with a Hamiltonian $H$ of the
form (\ref{Eq:SH}). (For two particles, this form of the Hamiltonian $H$
holds only for equal masses $m$ of the bound-state
constituents.)\end{enumerate}(For a more detailed account of the reduction of
the Bethe--Salpeter equation to the spinless Salpeter equation, consult,
e.g., the introductory sections of Refs.~\cite{Lucha98O,Lucha98D}.) This wave
equation describes the bound states of spin-zero particles (scalar bosons) as
well as the spin-averaged spectra of the bound states of fermions.

In this paper we consider potentials which are at the same time convex
transformations $V(r)=g(h(r))$ of the Coulomb potential $h(r)=-1/r$ and
concave transformations of the harmonic-oscillator potential $h(r)=r^2.$ The
reason for this is that spectral information is known for these two ``basis''
potentials $h(r)$. Thus the class of potentials is those $V(r)$ that have a
dual representation$$V(r)=g^{(1)}\left(-\frac{1}{r}\right)=g^{(2)}(r^{2})\
,$$in which $g^{(1)}$ is convex (${g^{(1)}}''>0$) and $g^{(2)}$ is concave
(${g^{(2)}}''<0$). An example of a potential in this class
is\begin{equation}V(r)=-\frac{c_1}{r}+c_2\ln r+c_3r+c_4r^2\
,\label{Eq:pot-class}\end{equation}where the coefficients $\{c_{i}\}$ are not
negative and are not all zero. Thus tangent lines to the transformation
function $g(h)$ are of the form $ah+b$ and are either Coulomb
potentials~lying below $V$, or harmonic-oscillator potentials lying above
$V.$ This geometrical idea is the basis for our approach to the spectral
problem posed by $H.$ We shall consider applications of this idea to the
(nonrelativistic) Schr\"odinger problem, the relativistic kinetic-energy
operator, and the full Salpeter Hamiltonian in Secs.~\ref{Sec:NRET},
\ref{Sec:RKE}, and \ref{Sec:RET}, respectively. We shall show that all our
upper and lower bounds on the eigenvalues of the semirelativistic Salpeter
Hamiltonian $H$ of Eq.~(\ref{Eq:SH}) can be expressed in the compact form
$$E\approx\min_{r>0}\left[\beta\sqrt{m^2+\frac{P^2}{r^2}}+V(r)\right],$$where
$P$ is a constant for each bound, and a sign of approximate equality is used
to indicate that, for definite convexity of $g(h),$ the envelope theory
yields lower bounds for convex $g(h)$ and upper bounds for concave $g(h).$
The main purpose of the present considerations is to establish the general
envelope formalism in terms of which such bounds can be proved, and to
determine the appropriate values of $P.$

It is fundamental to our method that we first know something about the
spectrum~of~$H$ in those cases where $V(r)$ is one of the basis potentials,
i.e., the Coulomb and the~harmonic oscillator. These two spectra are
discussed in Sec.~\ref{Sec:CHOP} below. In Sec.~\ref{Sec:FP} we look at the
example of the Coulomb-plus-linear potential.

\section{The Coulomb and harmonic-oscillator potentials}\label{Sec:CHOP}
\subsection{Scaling behaviour}Since the two basis potentials are both pure
powers, it is helpful first to determine what~can be learnt about the
corresponding eigenvalues by the use of standard scaling arguments. By
employing a wave function $\phi(cr)$ depending on a scale variable $c>0,$ we
find the following scaling rule for the eigenvalues corresponding to
attractive pure power potentials~$v\,{\rm sgn}(q)r^q.$ The
Hamiltonian$$H=\beta\sqrt{m^2+{\bf p}^2}+v\,{\rm sgn}(q)r^q$$has the (energy)
eigenvalues $E(v,\beta,m),$ where$$E(v,\beta,m)=\beta
mE\!\left(\frac{v}{\beta m^{1+q}},1,1\right),\quad q\ge -1\ .$$The scaling
behaviour described by the above formula allows us to consider the
one-particle, unit-mass special case $m=\beta=1$ initially, that is to say,
to work w.l.o.g.\ with the operator$$H=\sqrt{1+{\bf p}^2}+v\,{\rm
sgn}(q)r^{q}\ .$$

\subsection{Coulomb potential}\label{Subsec:CP}In the case of the Coulomb
potential $V(r)=-v/r$ it is well known \cite{Herbst77} that the Hamiltonian
$H$ has a Friedrichs extension provided the coupling constant $v$ is not too
large. Specifically, it is necessary in this case that $v$ is smaller than a
critical value $v_{\rm c}$ of the coupling constant:$$v<v_{\rm
c}=\frac{2}{\pi}\ .$$With this restriction, a lower bound to the bottom of
the spectrum is provided by Herbst's formula\begin{equation}
E_0\ge\sqrt{1-(\sigma v)^{2}}\ ,\quad\sigma\equiv\frac{\pi}{2}\
.\label{Eq:Herbst-bound}\end{equation}By comparing the spinless Salpeter
problem to the corresponding Klein--Gordon equation, Martin and Roy
\cite{Martin89} have shown that if the coupling constant is further
restricted by $v<\frac{1}{2},$ then an improved lower bound is provided by
the expression\begin{equation}E_0\ge\sqrt{\frac{1+\sqrt{1-4v^{2}}}{2}}\
,\quad v<\frac{1}{2}\ .\label{Eq:MR-bound}\end{equation}It turns out that our
lower-bound theory has a simpler form when the Coulomb eigenvalue bound has
the form of Eq.~(\ref{Eq:Herbst-bound}) rather than that of
Eq.~(\ref{Eq:MR-bound}). For this reason, we have derived from
Eq.~(\ref{Eq:MR-bound}), by rather elementary methods, a new family of
Coulomb bounds. To this~end, we begin with the ansatz
$$\sqrt{\frac{1+\sqrt{1-4v^{2}}}{2}}\ge\sqrt{1-(\sigma v)^{2}}$$and look for
conditions under which it becomes true. Since both sides are positive, we may
square the ansatz and rearrange to yield$$v^2\le\frac{\sigma^2-1}{\sigma^4}\
.$$Meanwhile from (\ref{Eq:MR-bound}) we must always satisfy $v<\frac{1}{2}.$
This establishes the inequality we'll~need,
namely,\begin{equation}E_0\ge\sqrt{1-(\sigma v)^2}\ ,\quad
v\le\frac{\sqrt{\sigma^2-1}}{\sigma^2}<\frac{1}{2}\
.\label{Eq:NLB}\end{equation}Examples are\begin{eqnarray*}&&\sigma^2=2\
,\quad v\le\frac{1}{2}\ ;\\[1ex]
&&\sigma^2=\frac{3}{2}\left(3-\sqrt{5}\right)\approx 1.145898\ ,\quad
v\le\frac{1}{3}\ ;\\[1ex]&&\sigma^2=8-4\sqrt{3}\approx 1.071797\ ,\quad
v\le\frac{1}{4}\ .\end{eqnarray*}All these (lower) bounds are slightly weaker
than the Martin--Roy bound (\ref{Eq:MR-bound}) but above the Herbst bound
(\ref{Eq:Herbst-bound}). We note that these functions of the coupling
constant $v$ are all monotone and {\it concave\/}.

\subsection{Harmonic-oscillator potential}\label{Subsec:HOP}In the case of
the harmonic-oscillator potential, i.e., $V(r)=vr^2,$ much more is
known~\cite{Lucha99Q,Lucha99A}. In momentum-space representation the operator
${\bf p}$ becomes a $c$-variable and thus, from the spectral point of view,
the Hamiltonian$$H=\sqrt{1+{\bf p}^2}+vr^2$$is equivalent to the
Schr\"odinger operator\begin{equation}H=-v\Delta+\sqrt{1+r^2}\
.\label{Eq:SHam-HO}\end{equation}Since the potential in this operator
increases without bound, we know \cite{Reed78} that the spectrum of this
operator is entirely discrete. We call its eigenvalues ${\cal E}_{n\ell}(v),$
$n=1,2,\dots,$ $\ell=0,1,\dots,$ where $n$ counts the radial states in each
angular-momentum subspace labelled by $\ell.$ In what follows we shall either
approximate the eigenvalues ${\cal E}_{n\ell}(v)$ analytically or presume
that they are known numerically. The concavity of nonrelativistic
Schr\"odinger energy eigenvalues has been discussed in
Refs.~\cite{Narnhofer75,Thirring90}. Theorem~2~of Ref.~\cite{Hall83}
establishes concavity for the ground state; the same proof can be applied to
states which are (1) in the subspace corresponding~to angular momentum $\ell$
and (2) orthogonal to the first $n-1$ exact energy eigenstates in this
subspace. This establishes concavity also for all the higher Schr\"odinger
energy eigenvalues. Thus the eigenvalues ${\cal E}_{n\ell}(v),$ regarded as
functions of the coupling parameter $v,$ are {\it concave\/}.

\subsection{The spectral comparison theorem}\label{Subsec:SCT}For the class
of interaction potentials given by (\ref{Eq:pot-class}) with the coefficient
of the Coulombic~term satisfying the constraint$$\lim_{r\to
0}r^2V'(r)<\frac{2\beta}{\pi}\ ,$$the semirelativistic Salpeter Hamiltonian
$H$ is bounded below and is essentially self-adjoint \cite{Herbst77}.
Consequently, the discrete spectrum of $H$ is characterized variationally
\cite{Reed78} and it follows immediately from this that, if we compare two
such Hamiltonians $H$ having the potentials $V^{(1)}(r)$ and $V^{(2)}(r),$
respectively, and we know that $V^{(1)}(r)<V^{(2)}(r),$ then we may conclude
that the corresponding discrete eigenvalues $E_{n\ell}$ satisfy the
inequalities $E_{n\ell}^{(1)}<E_{n\ell}^{(2)}.$ We shall refer to this
fundamental result as the ``spectral comparison theorem.'' In the more~common
case of nonrelativistic dynamics, i.e., for a (nonrelativistic) kinetic term
of the form $\beta{\bf p}^2/2m$ in the Hamiltonian $H$, a constraint similar
to the above would hold for the coefficient of a possible additional
(attractive) $-1/r^2$ term in the potential $V(r).$

\section{Envelope representations for Schr\"odinger
operators}\label{Sec:NRET}
\begin{figure}[ht]\vspace*{-2cm}\begin{center}
\psfig{figure=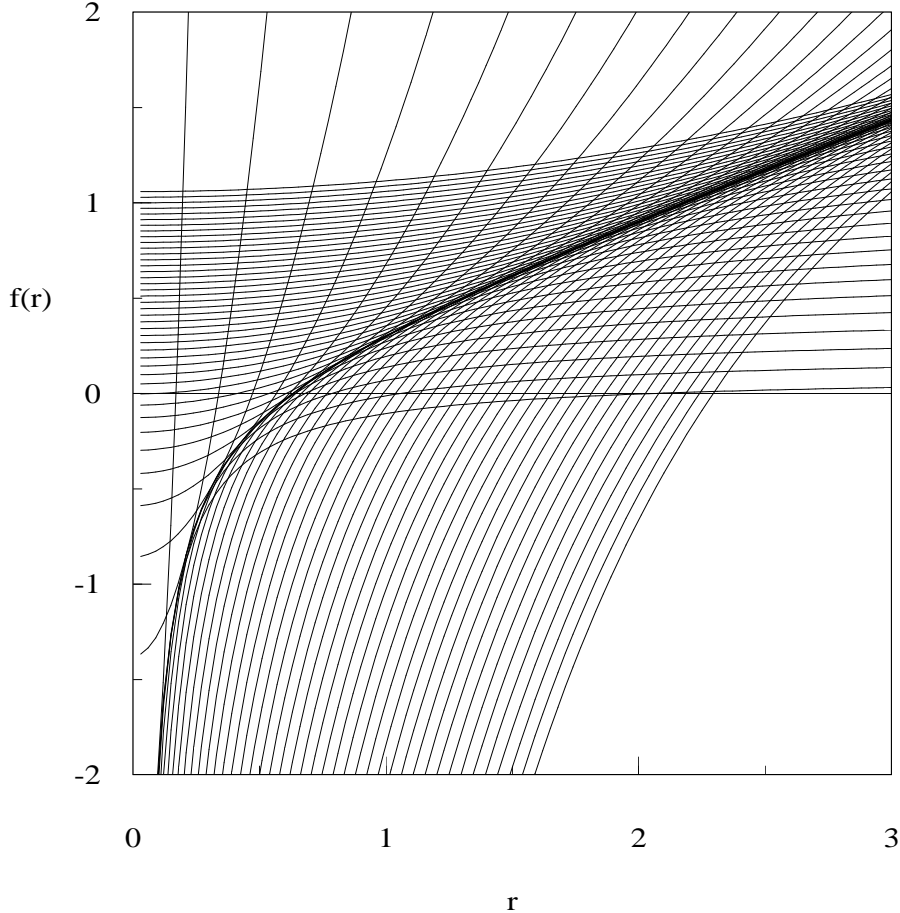,scale=.75}\vspace*{-1cm}\caption{The
``Coulomb-plus-linear'' potential shape $f(r)=-a/r+br$ represented as the
envelope curve of two distinct families of tangential potentials of the form
$\alpha h(r)+\beta.$ In the upper family $h(r)=r^2$ is the
harmonic-oscillator potential; in the lower family $h(r)=-1/r$ is~the Coulomb
potential. The adopted values of the relevant physical parameters $a$
and~$b$~are $a=0.2$ and $b=0.5.$}\label{Fig:shape}\end{center}\end{figure}

\begin{figure}[ht]\vspace*{-2cm}\begin{center}
\psfig{figure=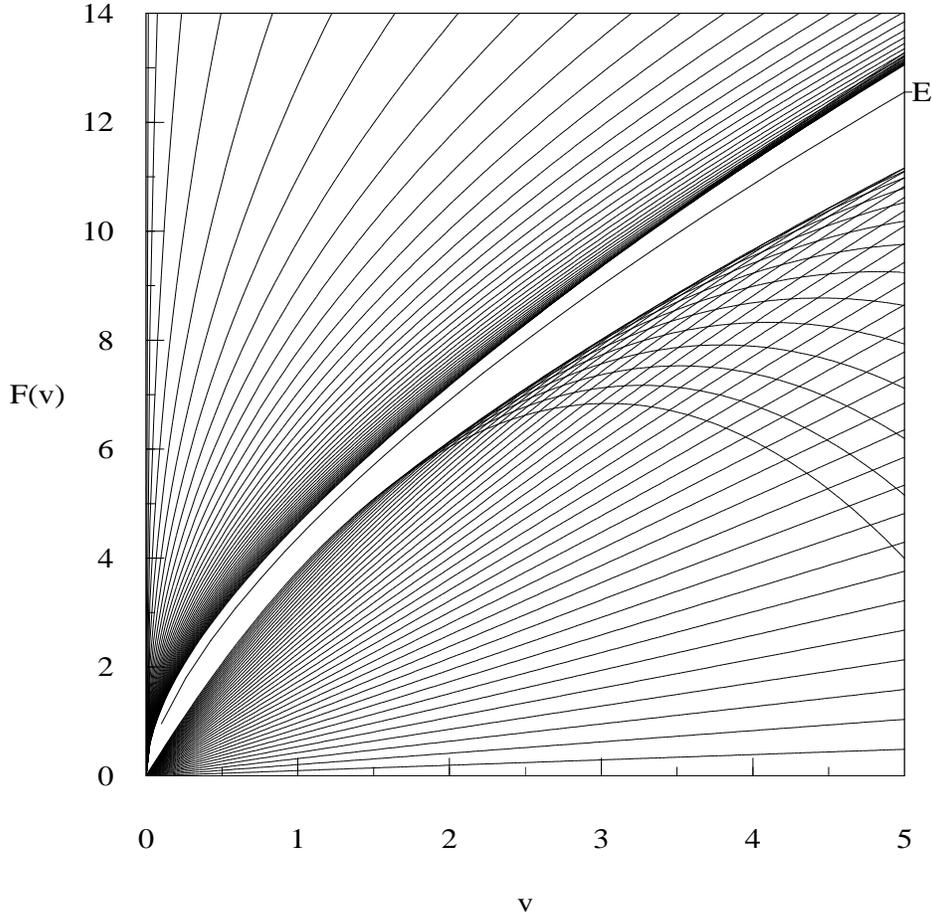,scale=.75}\vspace*{-1cm}\caption{The spectral
approximation corresponding to Fig.~\ref{Fig:shape}. Each ``tangential''
potential $f^{({\rm t})}(r)=\alpha h(r)+\beta$ generates a corresponding
tangential energy curve $F^{({\rm t})}(v)=H(\alpha v)+\beta v.$ The envelopes
of these spectral families generate upper and lower bounds to the exact curve
$E=F(v),$ shown here for the case $(n,\ell)=(2,4).$}
\label{Fig:bounds}\end{center}\end{figure}

We distinguish a potential $V(r)=vf(r)$ from its
shape $f(r),$ where the positive parameter~$v$ is often called the ``coupling
constant.'' The idea behind envelope representations \cite{Hall83,Hall84} is
suggested by the question: if one potential $f(r)$ can be written as a smooth
transformation $f(r)=g(h(r))$ of another potential $h(r),$ what spectral
relationship might this induce?~We consider potential shapes that support at
least one discrete eigenvalue for sufficiently large values of the coupling
$v$ and suppose for the sake of definiteness that the lowest eigenvalue~of
$-\Delta+vh(r)$ is given by $H(v)$ and that of $-\Delta+vf(r)$ by $F(v).$ If
the transformation function $g(h)$ is smooth, then each tangent to $g$ is an
affine transformation of the ``envelope basis''~$h$ of the form $f^{({\rm
t})}(r)=a(t)h(r)+b(t),$ where $r=t$ is the point of contact. The
coefficients~$a(t)$ and $b(t)$ are obtained by demanding that the
``tangential potential'' $f^{({\rm t})}(r)$ and its derivative agree with
$f(r)$ at the point of contact $r=t.$ Thus we have$$a(t)=\frac{f'(t)}{h'(t)}\
,\quad b(t)=f(t)-a(t)h(t)\ .$$The corresponding geometrical configuration is
illustrated in Fig.~\ref{Fig:shape} in which the potential~$f$ is chosen to
be the Coulomb-plus-linear potential, $f(r)=-1/r+r,$ and the envelope
basis~$h$ is, for the upper family, the harmonic-oscillator potential
$h(r)=r^2$ and, for the lower family, the Coulomb potential $h(r)=-1/r.$ The
spectral function $F^{({\rm t})}$ for the tangential potential $f^{({\rm
t})}(r)=a(t)h(r)+b(t)$ is given by $F^{({\rm t})}(v)=H(va(t))+vb(t).$ If the
transformation $g(h)$ has definite convexity, say $g''(h)>0,$ then each
tangential potential~$f^{({\rm t})}(r)$ lies beneath~$f(r)$ and, as a
consequence of the spectral comparison theorem, we know that each
corresponding tangential spectral function $F^{({\rm t})}(v),$ and the
envelope of this set, lie beneath $F(v).$ Similarly, in the case where $g$ is
concave, i.e., $g''(h)<0$, we obtain upper bounds to $F(v).$ These purely
geometrical arguments, depending on the spectral comparison theorem, extend
easily to~the excited states of the problem under consideration. The spectral
curves corresponding to~the envelope representations for the potential in
Fig.~\ref{Fig:shape} are shown in Fig.~\ref{Fig:bounds} for the excited state
$(n,\ell)=(2,4).$ For comparison the exact curve $E=F(v)$ is also shown in
Fig.~\ref{Fig:bounds}; this curve will be close to the Coulomb envelope for
large $v$ and to the oscillator envelope for~small~$v.$ Of course, the
envelopes of which we speak still have to be determined explicitly.
Extensions of this idea to completely new problems, such as simultaneous
transformations of each of~a number of potential terms \cite{Hall98}, the
Dirac equation \cite{Hall86,Hall99}, or the spinless-Salpeter problem of the
present paper, are best formulated initially with the basic argument outlined
above.

In the 1-term case the question arises as to whether there is a simple way of
determining the envelopes of the families of upper and lower spectral
functions. One effective solution of this problem is by the use of ``kinetic
potentials'' which were introduced \cite{Hall83,Hall84} precisely for this
purpose. The idea is as follows. To each spectral function $F_{n\ell}(v)$
there is a corresponding ``kinetic potential'' (that is, a {\it minimum mean
iso-kinetic potential\/}) $\bar f_{n\ell}(s).$ The relationship between
$F_{n\ell}$ and $\bar f_{n\ell}$ is invertible and is essentially that of a
Legendre transformation \cite{Gelfand}:~we can prove in general that $F$ is
concave, $\bar f(s)$ is convex, and$$F''(v)\bar f''(s)=-\frac{1}{v^3}\ .$$The
explicit transformation formulas are as follows:\begin{equation}\bar
f_{n\ell}(s)=F'_{n\ell}(v)\ ,\quad s=F_{n\ell}(v)-vF'_{n\ell}(v)
\label{eq:TF1}\end{equation}and\begin{equation}\frac{F_{n\ell}(v)}{v}=\bar
f_{n\ell}(s)-s\bar f'_{n\ell}(s)\ ,\quad\frac{1}{v}=-\bar f'_{n\ell}(s)\
.\label{eq:TF2}\end{equation}An {\it a-priori\/} definition of the
ground-state kinetic potential $\bar f_{10}(s)=\bar f(s)$ is given by$$\bar
f(s)=\inf_{{{\scriptstyle\psi\in{\cal
D}(H)}\atop{\scriptstyle(\psi,\psi)=1}}\atop{\scriptstyle(\psi,-\Delta\psi)=s}}
(\psi,f\psi)\ ,$$where ${\cal D}\subset L^{2}(\mbox{\vrule R}^3)$ is the
domain of the Hamiltonian. The definition for the excited states~is a little
more complicated \cite{Hall84} and, in view of (\ref{eq:TF1}) and
(\ref{eq:TF2}), will not be needed in what follows.

What is crucial is that the spectral functions, either exact or approximate,
are recovered from the corresponding kinetic potentials by a minimization
over the kinetic-energy variable $s.$ In this way the total minimization
required by the minimum--maximum principle \cite{Reed78,Thirring90} has been
divided into two steps: the first is constrained by $(\psi,-\Delta\psi)=s$
and the second~is~a minimization over $s.$ We have in all
cases:$$F_{n\ell}(v)=\min_{s>0}\left[s+v\bar f_{n\ell}(s)\right].$$Another
form of this expression is possible for the kinetic potential is monotone and
allows us to change variables $(s\rightarrow r)$ by $f(r)=\bar f_{n\ell}(s).$
Thus we have\begin{equation}
F_{n\ell}(v)=\min_{r>0}\left[K^{(f)}_{n\ell}(r)+vf(r)\right],\quad
K^{(f)}_{n\ell}=\bar f_{n\ell}^{-1}\circ f\ .\label{Eq:CV}\end{equation}The
two corresponding expressions of the envelope approximation then become$$\bar
f_{n\ell}(s)\approx g(\bar h_{n\ell}(s))\ ,\quad{\rm or}\quad
K^{(f)}_{n\ell}\approx K^{(h)}_{n\ell}\ .$$The second form (\ref{Eq:CV}) of
the explicit expression for $F_{n\ell}(v)$ isolates the potential
shape~$f$~itself and leads to an inversion sequence \cite{Hall00a} which
reconstructs the potential from a single given spectral function; but this is
another story \cite{Hall95,Hall99a}.

It is useful here to provide the formulas for the kinetic potentials $\bar f$
corresponding~to~pure power-law potentials $V(r)=v\,{\rm sgn}(q)r^q.$
Elementary scaling arguments for the Hamiltonian$$H=-\Delta+v\,{\rm
sgn}(q)r^q$$show that the eigenvalues satisfy
$$F_{n\ell}(v)=v^{\frac{2}{2+q}}F_{n\ell}(1)\ ,$$where
$F_{n\ell}(1)=E_{n\ell}(q)$ are the eigenvalues of $H$ with coupling $v=1,$
i.e., of $-\Delta+{\rm sgn}(q)r^q.$ From Eq.~(\ref{eq:TF1}) we immediately
find that the kinetic potentials $\bar f_{n\ell}(s)$ for these
potentials~$V(r)$ are~given by\begin{equation}\bar f_{n\ell}(s)=\frac{2}{q}
\left|\frac{qE_{n\ell}(q)}{2+q}\right|^{\frac{2+q}{2}}s^{-\frac{q}{2}}\ .
\label{Eq:PL-KP}\end{equation}Meanwhile the corresponding $K$-functions all
have the same simple form \cite{Hall93}$$K_{n\ell}^{(q)}(r)=
\left(\frac{P_{n\ell}(q)}{r}\right)^2\ ,$$where the $P$ numbers are given by
$$P_{n\ell}(q)=\left|E_{n\ell}(q)\right|^{\frac{2+q}{2q}}
\left(\frac{2}{2+q}\right)^{\frac{1}{q}}
\left|\frac{q}{2+q}\right|^{\frac{1}{2}}\ ,\quad q\ne 0\ .$$Consequently, the
power-law kinetic potentials (\ref{Eq:PL-KP}) may be expressed in the simple
form$$\bar f_{n\ell}(s)={\rm
sgn}(q)\left(\frac{P_{n\ell}(q)}{\sqrt{s}}\right)^q\ ,\quad q\ne 0\ .$$Some
of the eigenvalues $E,$ and thus the corresponding $P$ numbers, are known
exactly from elementary quantum mechanics. From the known eigenvalues $E$ for
the Coulomb potential, $E_{n\ell}(-1)=-[2(n+\ell)]^{-2},$ and the
harmonic-oscillator potential, $E_{n\ell}(2)=4n+2\ell-1,$ we immediately
obtain the corresponding $P$ numbers:\begin{eqnarray*}P_{n\ell}(-1)&=&n+\ell\
,\\[1ex]P_{n\ell}(2)&=&2n+\ell-\frac{1}{2}\ .\end{eqnarray*}The case $q=0$
corresponds {\it exactly\/} to the logarithmic potential
\cite{Quigg79,Hall00b}; the $P$ numbers~for the lowest states of the
logarithmic and the linear potentials may be found in
Table~\ref{Tab:P-numbers}.

\begin{table}[ht]\caption{Numerical values for the $P$ numbers for the
logarithmic potential ($q=0$) and the linear potential ($q=1$) used in the
Schr\"odinger eigenvalue formula (\ref{Eq:EVs}).}\label{Tab:P-numbers}\small
\begin{center}\begin{tabular}{ccrr}\hline\hline&&\\[-1.5ex]
\multicolumn{1}{c}{$n$}&\multicolumn{1}{c}{$\ell$}&
\multicolumn{1}{c}{$P_{n\ell}(0)$}&\multicolumn{1}{c}{$P_{n\ell}(1)$}\\[1ex]
\hline\\[-1.5ex]
1&0&1.21867&1.37608\\2&0&2.72065&3.18131\\3&0&4.23356&4.99255\\
4&0&5.74962&6.80514\\5&0&7.26708&8.61823\\[.5ex]
1&1&2.21348&2.37192\\2&1&3.68538&4.15501\\3&1&5.17774&5.95300\\
4&1&6.67936&7.75701\\5&1&8.18607&9.56408\\[.5ex]
1&2&3.21149&3.37018\\2&2&4.66860&5.14135\\3&2&6.14672&6.92911\\
4&2&7.63639&8.72515\\5&2&9.13319&10.52596\\[.5ex]
1&3&4.21044&4.36923\\2&3&5.65879&6.13298\\3&3&7.12686&7.91304\\
4&3&8.60714&9.70236\\5&3&10.09555&11.49748\\[.5ex]
1&4&5.20980&5.36863\\2&4&6.65235&7.12732\\3&4&8.11305&8.90148\\
4&4&9.58587&10.68521\\5&4&11.06725&12.47532\\[.5ex]
1&5&6.20936&6.36822\\2&5&7.64780&8.12324\\3&5&9.10288&9.89276\\
4&5&10.56970&11.67183\\5&5&12.04517&13.45756\\[1ex]
\hline\hline\end{tabular}\end{center}\end{table}\normalsize

In summary, if the potential $V(r)$ is a smooth transformation $V(r)=g({\rm
sgn}(q)r^q)$ of~the pure power-law potential ${\rm sgn}(q)r^q$, then the
eigenvalues of$$H=-\Delta+V(r)$$are given approximately by the
expression\begin{equation}E_{n\ell}\approx\min_{r>0}
\left[\frac{P_{n\ell}^2(q)}{r^2}+V(r)\right].\label{Eq:EVs}\end{equation}
Here, a sign of approximate equality is used to indicate that, for a definite
convexity~of~$g(h),$ Eq.~(\ref{Eq:EVs}) yields lower bounds for convex $g$
($g''>0$) and upper bounds for concave $g$ ($g''<0$). The numbers
$P_{n\ell}(q)$ can be derived from the eigenvalues of the operator
$-\Delta+{\rm sgn}(q)r^q.$

The lower bounds derived in the framework of envelope theory can be improved
by~use~of the refined comparison theorems of Ref.~\cite{Hall92}, which allow
comparison potentials~to intersect; a detailed study of the latter approach
is, however, beyond the scope of the present~analysis.

As an immediate application we consider the (nonrelativistic) Schr\"odinger
Hamiltonian (\ref{Eq:SHam-HO}) for the semirelativistic spinless-Salpeter
harmonic-oscillator problem (\ref{Eq:SH}). Here we have$$H=-v\Delta+V(r)\
,$$with$$V(r)=\beta\sqrt{m^2+r^2}\ ;$$this potential is a convex
transformation of a linear potential and a concave transformation of a
harmonic-oscillator potential. We conclude therefore from Eq.~(\ref{Eq:EVs})
(see also Ref.~\cite{Lucha00-HO}):\begin{equation}
\min_{r>0}\left[v\frac{P_{n\ell}^2(1)}{r^2}+\beta\sqrt{m^2+r^2}\right]\leq
{\cal E}_{n\ell}(v)\leq
\min_{r>0}\left[v\frac{P_{n\ell}^2(2)}{r^2}+\beta\sqrt{m^2+r^2}\right],
\label{Eq:SEVF-HO}\end{equation}where the numbers $P_{n\ell}(1)$ are given in
Table~\ref{Tab:P-numbers} and $P_{n\ell}(2)=2n+\ell-\frac{1}{2}$. By a simple
change of variables, $r\rightarrow r'=P/r,$ we are able to recast the
inequalities (\ref{Eq:SEVF-HO}) into the ``preferred'' form$${\cal
E}_{n\ell}(v)\approx
\min_{r>0}\left[\beta\sqrt{m^2+\frac{P^2_{n\ell}}{r^2}}+V(r)\right],\quad
V(r)=vr^2\ ,$$in which the function to be minimized is simply the
spinless-Salpeter Hamiltonian with~the momentum operator $|{\bf p}|$ replaced
by $P/r;$ the $P$ numbers yielding upper and lower bounds are as in
Eq.~(\ref{Eq:SEVF-HO}). Interestingly, the upper and lower bounds in
Eq.~(\ref{Eq:SEVF-HO}) are equivalent~to~the corresponding bounds
obtained~in~Ref.~\cite{Lucha99A}; however, these earlier specific bounds were
not derived as part of the general envelope theory. If we approximate the
square root from~above {\it again\/}, by using the elementary
inequality$$\sqrt{m^2+r^2}\le m+\frac{r^2}{2m}\ ,$$we obtain from
(\ref{Eq:SEVF-HO}) the weaker upper bound$${\cal E}_{n\ell}(v)\le\beta
m+\sqrt{\frac{\beta v}{2m}}(4n+2\ell-1)\ ,$$which is identical to that given
by the general ``Schr\"odinger upper~bound''
\cite{Lucha96a,Lucha98O,Lucha98D} obtained by initially approximating the
relativistic kinetic-energy operator above by$$\beta\sqrt{m^2+{\bf
p}^2}\le\beta\left(m+\frac{{\bf p}^2}{2m}\right).$$The latter upper bound,
and an improvement on it, will be discussed in the next section.

\section{Envelope approximations for the relativistic kinetic
energy}\label{Sec:RKE}The relativistic kinetic-energy operator
$T=\beta\sqrt{m^2+{\bf p}^2}$ is a concave transformation of ${\bf p}^2.$
Thus ``tangent lines'' to this operator all are of the form $a{\bf p}^2+b$
and each one generates a Schr\"odinger operator that provides an upper bound
to $T.$ In a given application with given parameter values, one can search
for the best such upper bound. By elementary analysis~we can establish the
operator inequality\begin{equation}H=\beta\sqrt{m^2+{\bf p}^2}+V(r)\le
\frac{\beta}{2}\left(-\frac{\Delta}{\mu}+\mu+\frac{m^2}{\mu}\right)+V(r)\
,\label{Eq:OI}\end{equation}where $\mu=\sqrt{m^2+p_1^2},$ and $|{\bf p}|=p_1$
is the ``point of contact'' of the tangent line with the square-root
function. The inequality (\ref{Eq:OI}) is identical to that obtained
\cite{Lucha96a} by employing the inequality $(T-\mu)^2\ge 0.$ Optimization
over $\mu$ for the Coulomb case $V(r)=-v/r$ recovers the explicit upper-bound
formula of Ref.~\cite{Lucha96a}:$$E_{n\ell}(v)\le
m\beta\sqrt{1-\left(\frac{v}{\beta(n+\ell)}\right)^2}\ .$$In the case of the
harmonic-oscillator potential, $V(r)=vr^2,$ we obtain the upper
bounds~\cite{Lucha99A}$$E_{n\ell}(v)\le\min_{\mu>0}\left[\sqrt{\frac{\beta
v}{2\mu}}(4n+2\ell-1)+\frac{\beta}{2}\left(\mu+\frac{m^2}{\mu}\right)\right].$$
(Brief reviews of analytical upper bounds on the energy eigenvalues of the
spinless Salpeter equation derived by combining operator inequalities with
the minimum--maximum principle may be found in
Refs.~\cite{Lucha98O,Lucha98D,Lucha98R}.)

The strategy of the present section is to regard the relativistic
kinetic-energy operator~$T$ as a concave function of ${\bf p}^2,$ so that
``tangent lines'' generate ``upper'' Schr\"odinger operators. This general
approach leads to the same upper bounds as those of Martin \cite{Martin88}
who used~the particular square-root form of the relativistic kinetic energy
to construct an operator whose positivity yields the bounds.

\section{Envelope approximations for Salpeter Hamiltonians}\label{Sec:RET}
\subsection{The principal envelope formula}Let us now turn to our main topic
and consider the spinless-Salpeter Hamiltonian of Eq.~(\ref{Eq:SH}),
$$H=\beta\sqrt{m^2+{\bf p}^2}+V(r)\ ,$$and its eigenvalues $E.$ We shall
assume that the potential $V(r)$ is a smooth transformation $V(r)=g(h(r))$ of
another potential $h(r)$ and that $g$ has definite convexity so that
we~obtain bounds to the energy eigenvalues $E$. We suppose that the ``basis''
potential $h(r)$ generates~a ``tangential'' Salpeter problem$${\cal
H}=\beta\sqrt{m^2+{\bf p}^2}+vh(r)\ ,$$for which the eigenvalues $e(v),$ or
bounds to them, are known. We shall follow here as~closely as possible the
development in Sec.~\ref{Sec:NRET} for the corresponding Schr\"odinger
problem.

First of all, we recall that the approximations or bounds to the energy
eigenvalues of~the relativistic Coulomb and harmonic-oscillator problems we
shall eventually use from Secs.~\ref{Sec:CHOP} and \ref{Sec:NRET}, when
regarded as functions of the coupling~$v,$ are all {\it concave\/}.
Furthermore, it is~easy to convince oneself that all the (unknown) energy
functions $e(v)$ of the ``tangential'' Salpeter problem~are {\it concave\/},
that is, $e''(v)<0.$ Suppose that the exact eigenvalue and~(normalized)
eigenvector for the problem posed by the ``tangential'' Hamiltonian$${\cal
H}=\beta\sqrt{m^2+{\bf p}^2}+vh(r)$$are $e(v)$ and $\psi(v,r).$ Then, by
differentiating the expectation value $(\psi,{\cal H}\psi)$ with respect~to
the coupling $v,$ we find$$e'(v)=(\psi,h\psi)\ .$$If we now apply $\psi(v,r)$
as a trial vector to estimate the energy of the operator$$\beta\sqrt{m^2+{\bf
p}^2}+uh(r)\ ,$$in which $v$ has been replaced by $u,$ we obtain an upper
bound to the corresponding energy function $e(u)$ which may be written in the
form$$e(u)\leq e(v)+(u-v)e'(v)\ .$$This inequality tells~us that the function
$e(u)$ lies beneath its tangents; that is to say,~$e(u)$~is {\it concave\/}.
Convexity properties of the energy functions of the corresponding
(nonrelativistic) Schr\"odinger problem have been investigated in
Refs.~\cite{Narnhofer75,Hall83,Thirring90}.

Next, in order to prove the main result of this section, the ``principal
envelope formula,'' we begin by using an envelope representation for the
potential $V(r)$ in the Hamiltonian~(\ref{Eq:SH}) and then demonstrate that
all the spectral formulas that follow possess a certain structure. Finally,
as an application, we specialize to the case of pure power-law ``basis''
potentials~$h(r)$ and, more particularly, to the Coulomb potential and the
harmonic-oscillator potential for which, at this time, we have spectral
information [cf.\ the discussions in Secs.~\ref{Subsec:CP} and
\ref{Subsec:HOP}, and the exact bounds (\ref{Eq:SEVF-HO}) on the energy
levels ${\cal E}_{n\ell}(v)$ of the relativistic harmonic oscillator].

The tangential potentials we shall employ have the form $V^{({\rm
t})}(r)=a(t)h(r)+b(t),$~where, as in the Schr\"odinger case, the coefficients
$a(t)$ and $b(t)$ are given by$$a(t)=\frac{V'(t)}{h'(t)}=g'(h(t))\ ,\quad
b(t)=V(t)-a(t)h(t)=g(h(t))-g'(h(t))h(t)\ ,$$and $r=t$ is the point of contact
of the potential $V(r)$ and its tangent $V^{({\rm t})}(r)$. If, for~the~sake
of definiteness, we assume that $V=g(h)$ with $g$ concave (i.e., $g''<0$), we
obtain a family~of upper bounds given by$$E\le\varepsilon(t)=e(a(t))+b(t)\
.$$The best of these is given by optimizing over $t$:$$E\le\varepsilon(\hat
t)=e(a(\hat t))+b(\hat t)\ ,$$where $\hat t,$ the value of $t$ which
optimizes these bounds, is to be determined as the
solution~of$$e'(g'(h(\hat{t})))=h(\hat{t})\ .$$In the spirit of the Legendre
transformation \cite{Gelfand} we now consider another problem which~has the
same solution; this second problem is the one that provides us with our
basic~eigenvalue formula. We consider$${\cal
E}\equiv\min_{v>0}[e(v)-ve'(v)+g(e'(v))]\ ,$$which is well defined since
$e(v)$ is concave. The solution has the critical point$$\hat v=g'(e'(\hat
v))\ .$$If we now apply the correspondence $h(\hat t)=e'(\hat{v}),$ it
follows that the critical point $\hat v$ becomes$$\hat v=g'(h(\hat t))\
,$$and the tangential-potential coefficients $a$ and $b$
become\begin{equation}a(\hat t)=g'(e'(v))=v\ ,\quad b(\hat
t)=g(e'(v))-ve'(v)\ ,\quad v=\hat v\ .\label{Eq:TPC}\end{equation}Meanwhile
the original critical (energy) value is given by$$\varepsilon(\hat
t)=e(a(\hat t))+b(\hat t)=e(v)-ve'(v)+g(e'(v))\ ,\quad v=\hat v\ .$$Thus we
conclude that the spectral approximation obtained by envelope methods is
given by the following ``principal envelope formula:''\begin{equation}
E\approx{\cal E}\equiv\min_{v>0}[e(v)-ve'(v)+g(e'(v))]\
.\label{Eq:PEF}\end{equation}If $g$ is concave (that is, $g''<0$), then
$E\le{\cal E};$ if $g$ is convex (that is, $g''>0$), then $E\ge{\cal E}.$
From the above considerations it follows immediately that, if the {\it
exact\/} energy function $e(v)$ corresponding to the basis potential $h$ is
not available, then, for $g(h)$ concave, concave~{\it upper\/} approximations
$e_{\rm u}(v)>e(v)$ or, for $g(h)$ convex, concave {\it lower\/}
approximations $e_{\rm l}(v)<e(v)$ may be used instead of the exact energy
function $e(v)$ in the principal envelope formula~(\ref{Eq:PEF}). Then all
the lower tangents will lie even lower and all the upper tangents will lie
even~higher. If $g$ is convex, we obtain a lower bound; if $g$ is concave, we
obtain an upper bound; because~of the concavity of $e(v),$ {\it this\/}
extremum is a minimum in {\it both\/} cases. If we wish to use numerical
solutions to the ``basis'' problem (generated by $h(r)$), or if a completely
new energy-bound expression becomes available, the principal envelope formula
(\ref{Eq:PEF}) is the relation that would at first be used.

Interestingly, in the formula (\ref{Eq:PEF}) the tangential-potential
apparatus is no longer evident; only the correct convexity is required. As in
the Schr\"odinger case \cite{Hall83}, once we have the basic result, the
reformulation in terms of ``kinetic potentials'' is often useful: the kinetic
potential $\bar h(s)$ corresponding to some basis potential $h(r)$ is given
by the Legendre transformation~\cite{Gelfand}$$\bar h(s)=e'(v)\ ,\quad
s=e(v)-ve'(v)\ .$$Meanwhile the envelope approximation has the
kinetic-potential expression $\bar{V}(s)\approx g(\bar h(s)).$

For both the Coulomb lower bounds (\ref{Eq:Herbst-bound}) or (\ref{Eq:NLB})
and the harmonic-oscillator upper bounds (\ref{Eq:SEVF-HO}) which we have at
present, we may express our general results in a special common~form which
will now be derived.

\subsection{The Coulomb lower bound}We consider first the Coulomb lower bound
in which we assume that the potential $V(r)$~is~a convex transformation
$V(r)=g(h(r))$ of the Coulomb potential $h(r)=-1/r.$ According~to
Sec.~\ref{Subsec:CP}, in this case all the ``lower'' $e_{\rm l}(v)$ have been
arranged---with the parameters $\beta$ and~$m$ returned---in the form$$e_{\rm
l}(v)=\beta m\sqrt{1-\left(\frac{\sigma v}{\beta}\right)^2}\ .$$From this it
follows by elementary algebra that if we define a new optimization
variable~$r$~by $e_{\rm l}'(v)=h(r)=-1/r,$ we have$$e_{\rm l}(v)-ve_{\rm
l}'(v)=\beta\sqrt{m^2+\frac{P^2}{r^2}}\ ,\quad P\equiv\frac{1}{\sigma}\ .$$
Consequently, the lower bound on the energy eigenvalues $E$ of the spinless
Salpeter equation becomes\begin{equation}
E\ge\min_{r>0}\left[\beta\sqrt{m^2+\frac{P^2}{r^2}}+V(r)\right],\quad v<\beta
v_P\ .\label{Eq:SSE-LB}\end{equation}Here the boundary value $v_P$ of the
Coulomb coupling $v$ is given, when simply determined by the requirement of
boundedness from below of the operator (\ref{Eq:SH}), by the critical
coupling~$v_{\rm c}$,$$v_P=v_{\rm c}=\frac{2}{\pi}\ ,$$and, when arising from
the region of validity of our Coulomb-like family of lower
bounds~(\ref{Eq:NLB}), via $P=1/\sigma,$ by\begin{equation}
v_P=P\sqrt{1-P^2}<\frac{1}{2}\ .\label{Eq:CCUBvP}\end{equation}Some
$\{P,v_P\}$ pairs may be found in Table~\ref{Tab:PvP-pairs}; others can be
easily generated from the upper bound on the coupling $v$ given in
Eq.~(\ref{Eq:CCUBvP}). The meaning of the Coulomb-coupling~constraint is
$a(\hat t)<\beta v_P,$ where $a$ is the coefficient in the tangential Coulomb
potential given by (\ref{Eq:TPC}).

\begin{table}[ht]\caption{Explicit values of some $\{P,v_P\}$ pairs, obtained
via the equality $P=1/\sigma$ either~from the Herbst lower bound
(\ref{Eq:Herbst-bound}) or the expression (\ref{Eq:NLB}) for our new lower
bounds on the spectrum of the spinless relativistic Coulomb problem (in three
spatial dimensions).}\label{Tab:PvP-pairs}\small
\begin{center}\begin{tabular}{cc}\hline\hline\\[-1.5ex]
\multicolumn{1}{c}{$P$}&\multicolumn{1}{c}{$v_P$}\\[1ex]\hline\\[-1.5ex]
$\displaystyle\frac{2}{\pi}$&$\displaystyle\frac{2}{\pi}$\\[2ex]
$\displaystyle\frac{1}{\sqrt{2}}$&$\displaystyle\frac{1}{2}$\\[2ex]
$\sqrt{\displaystyle\frac{2}{9-3\sqrt{5}}}$&$\displaystyle\frac{1}{3}$\\[2ex]
$\displaystyle\frac{1}{2\sqrt{2-\sqrt{3}}}$&$\displaystyle\frac{1}{4}$\\[3ex]
\hline\hline\end{tabular}\end{center}\end{table}\normalsize

As a rather trivial consistency check of our formalism, the Coulomb lower
energy bound of Eq.~(\ref{Eq:SSE-LB}) may be applied to the Coulomb potential
$V(r)=-v/r$ in order to re-derive,~for $P=2/\pi,$ the Herbst formula
(\ref{Eq:Herbst-bound})---which is nothing else~but the starting point of the
present ``lower-bound'' considerations.

\subsection{The harmonic-oscillator upper bounds}Next, let us turn to the
harmonic-oscillator upper bounds. Our main assumption is here that
$V(r)=g(r^2),$ with $g''<0.$ In this case the only difficulty is that the
basis problem~$h(r)=r^2$ is equivalent to a Schr\"odinger problem whose
solution ${\cal E}_{n\ell}(v)$ is not known exactly. Following the discussion
of suitable bounds after the proof of the principal envelope formula,
Eq.~(\ref{Eq:PEF}), let us call the upper bound provided by
Eq.~(\ref{Eq:SEVF-HO}) $e_{\rm u}(v),$ and let us introduce the shorthand
notation $P_{n\ell}(2)=2n+\ell-\frac{1}{2}=P.$ Then we have~the following
parametric equations~for~$e_{\rm u}(v)$:$$e_{\rm
u}(v)=v\frac{P^2}{r^2}+\beta\sqrt{m^2+r^2}\ ,\quad v=\frac{\beta
r^4}{2P^2\sqrt{m^2+r^2}}\ ,\quad e_{\rm u}'(v)=\frac{P^2}{r^2}\ .$$By
substituting these expressions into the fundamental envelope formula
(\ref{Eq:PEF}) we obtain~the following upper bound on all the eigenvalues of
the spinless-Salpeter problem with potential $V(r)=g(r^2)$ and $g''<0$:
\begin{equation}
E_{n\ell}\le\min_{r>0}\left[\beta\sqrt{m^2+\frac{P^2}{r^2}}+V(r)\right],\quad
P=P_{n\ell}(2)=2n+\ell-\frac{1}{2}\ .\label{Eq:SSE-UB}\end{equation}

\section{The Coulomb-plus-linear (or ``funnel'') potential}\label{Sec:FP}In
order to illustrate the above results by a physically motivated example, let
us apply~these considerations to the Coulomb-plus-linear or (in view of its
shape) ``funnel'' potential$$V(r)=-\frac{c_1}{r}+c_2r\ ,\quad c_1\ge 0\
,\quad c_2\ge 0\ .$$(This potential provides a reasonable overall description
of the strong interactions of quarks in hadrons. For the phenomenological
description of hadrons in terms of both nonrelativistic and semirelativistic
potential models, see, e.g., Refs.~\cite{Lucha91:BSQ,Lucha92:QAQBS}.) By
choosing as basis potential the Coulomb potential $h(r)=-1/r,$ we may write
$V(r)=g(h(r))$ with$$g(h)=c_1h-\frac{c_2}{h}\ ,$$which is clearly a convex
function of $h<0$: $g''>0.$ Thus the convexity condition is~satisfied.
However, we are not free to choose the coupling constants $c_1$ and $c_2$ as
large as we please.~It is immediately obvious that, for a particular $\{P,
v_P\}$ pair, we must in any case have~$c_1<\beta v_P.$ For the full problem
the coefficient $c_2$ of the linear term will also be involved. The
coupling~$v$ we are concerned about is given by (\ref{Eq:TPC}). We
have$$v=g'(e'(v))=\frac{\beta
P^2}{r\sqrt{m^2+\left(\displaystyle\frac{P}{r}\right)^2}}
=c_1+\frac{c_2}{h^2}=c_1+c_2r^2\ .$$From this we obtain, for given values of
the parameters $m$ and $\beta$ and for a given $\{P, v_P\}$~pair, as a
sufficient condition for $v<\beta v_P$ the ``Coulomb coupling constant
constraint'' on the~two coupling strengths $c_1$ and $c_2$ in the funnel
potential:\begin{equation}c_1+\frac{P^2}{m^2}\left(\frac{P^2}{v_P^2}-1\right)
c_2<\beta v_P\ .\label{Eq:CCCC}\end{equation}In the case $\{P=1/\sqrt{2},
v_P=1/2\}$ and $\beta=m=1$ this condition reduces to
$c_1+\frac{1}{2}c_2<\frac{1}{2}.$ For Herbst's lower bound
(\ref{Eq:Herbst-bound}), i.e., $P=v_P=v_{\rm c}=2/\pi,$ this constraint
clearly yields~$c_1<\beta v_P.$ There is no escaping this feature of all
energy bounds involving the Coulomb potential: the constraint derives from
the fundamental observation that the Coulomb coupling $v$ must~not be too
large, so that the (relativistic) kinetic energy is able to counterbalance
the Coulomb potential in order to maintain the Hamiltonian (\ref{Eq:SH}) with
$V(r)=-v/r$ bounded from below.

For example, if we seek the largest allowed value of the parameter $P$ by
solving Eqs.~(\ref{Eq:CCUBvP}) and (\ref{Eq:CCCC}) together, we find that
this largest $P$ is given by
\begin{equation}\frac{c_2\sin^4t}{\cos^2t(\beta\sin t\cos t-c_1)}=m^2\ ,\quad
P\equiv\sin t\ .\label{Eq:P(m)}\end{equation}

\begin{figure}[ht]\vspace*{-2cm}\begin{center}
\psfig{figure=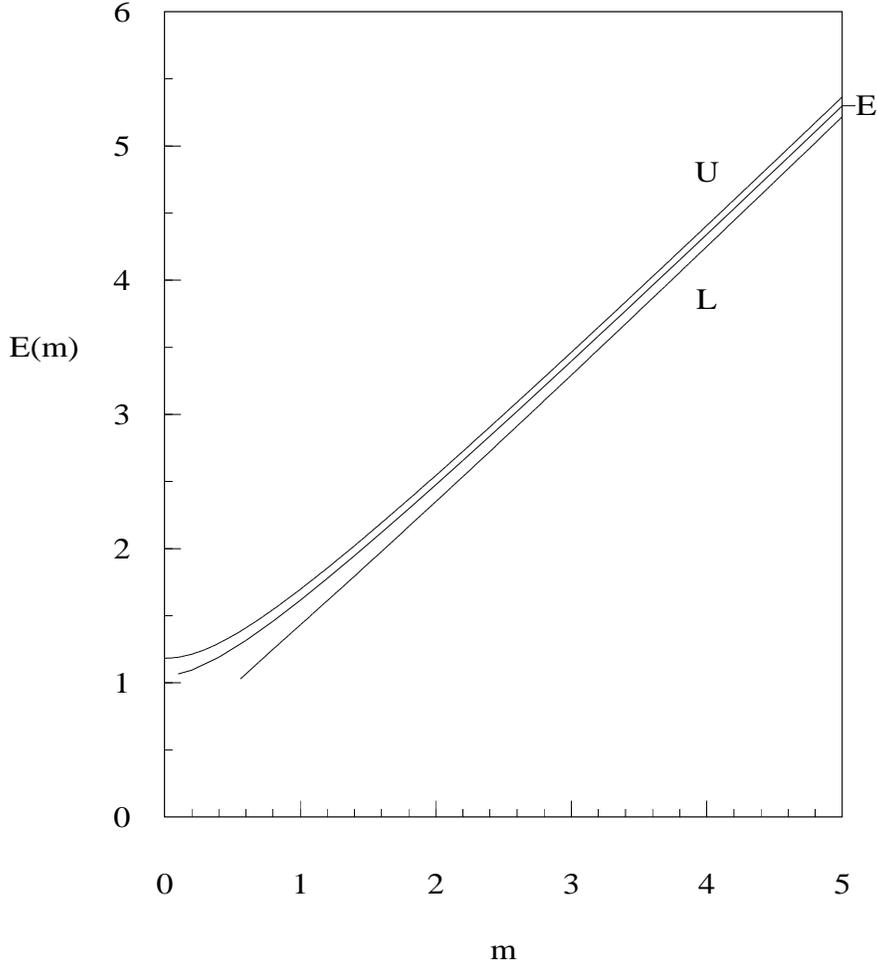,scale=0.792}\vspace*{-0.5cm}\caption{Lower bounds
(L), according to (\ref{Eq:SSE-LB}), and upper bounds (U), according to
(\ref{Eq:SSE-UB}),~on the energy eigenvalue $E$ of the ground state
[$(n,\ell)=(1,0)$] of the spinless Salpeter equation with a
Coulomb-plus-linear potential $V(r)=-c_1/r+c_2r,$ for $\beta=1,$ $c_1=0.1,$
and $c_2=0.25.$ The lower bound is given by the general result
(\ref{Eq:SSE-LB}) with the ``best'' $P(m)$ provided~by~(\ref{Eq:P(m)}). In
order to satisfy the Coulomb coupling constraint (\ref{Eq:P(m)}), the mass
$m$ must fulfil $m>\sqrt{5}/4.$ For comparison, a (very accurate)
Rayleigh--Ritz variational upper bound $E$ is
depicted~too.}\label{Fig:Bounds}\end{center}\end{figure}

For the Coulomb-plus-linear potential $V(r)=-c_1/r+c_2r$ under consideration,
Fig.~\ref{Fig:Bounds} shows the lower and upper bounds on the lowest energy
eigenvalue $E$ of the spinless Salpeter equation, given by the envelopes of
the lower and upper families of tangential energy curves (\ref{Eq:SSE-LB})
and (\ref{Eq:SSE-UB}), as functions $E(m)$ of the mass $m$ entering in the
semirelativistic Hamiltonian. In the case of the Coulomb lower bound
(\ref{Eq:SSE-LB}), we have employed for each $m$ the best possible $P(m)$
provided by (\ref{Eq:P(m)}). As $m\to 0,$ the ``basis'' Coulomb problem
$H=\beta\sqrt{m^2+{\bf p}^2}-v/r$~has energy $e(m)\to 0;$ thus the Coulomb
lower bound for a non-Coulomb problem becomes very weak for small values of
$m.$ Of course, Eq.~(\ref{Eq:SSE-UB}) provides us with rigorous upper bounds
for {\it every\/} energy level.

In order to get an idea of the location of the exact energy eigenvalues $E$,
Fig.~\ref{Fig:Bounds} also shows the ground-state energy curve $E(m)$
obtained by the Rayleigh--Ritz variational technique~\cite{Reed78} with the
Laguerre basis states for the trial space defined in Ref.~\cite{Lucha97}.
Strictly speaking, this energy curve represents only an upper bound to the
precise eigenvalue $E$. However,~from~the findings of Ref.~\cite{Lucha97} the
deviations of these Laguerre bounds from the exact eigenvalues~may be
estimated, for the superposition of 25 basis functions used here, to be of
the order of~1\,\%.

\begin{figure}[ht]\vspace*{-2cm}\begin{center}
\psfig{figure=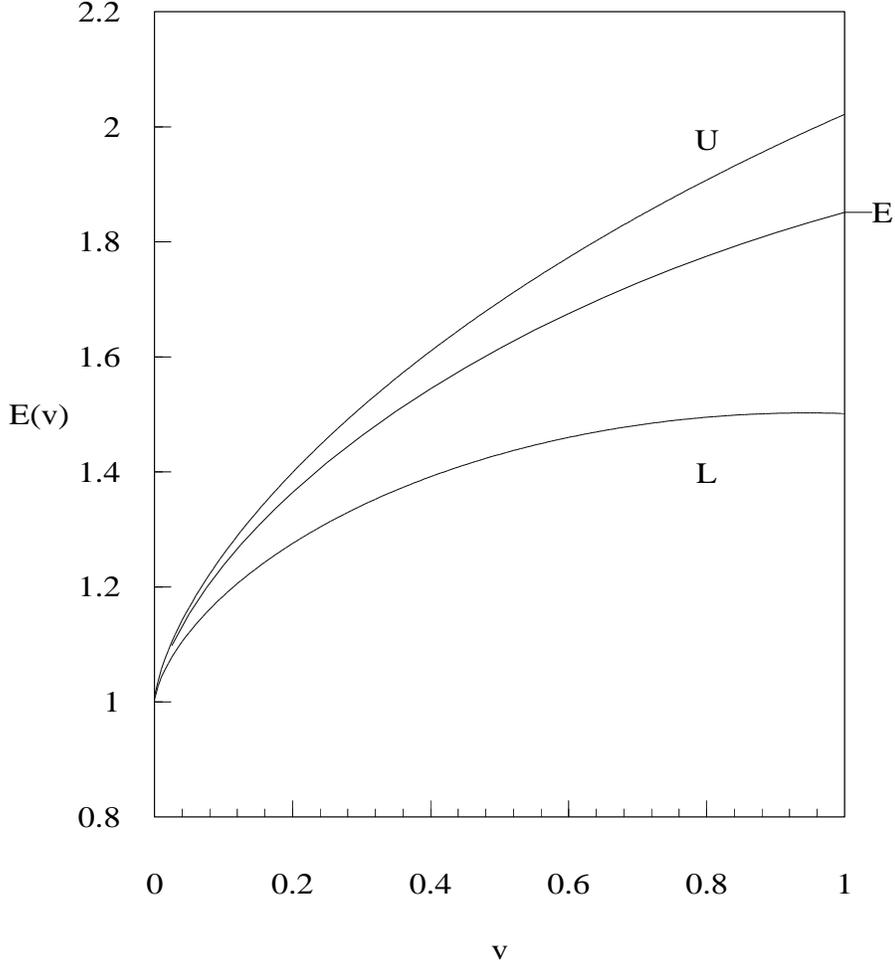,scale=0.792}\vspace*{-1cm}\caption{Lower bounds (L),
according to (\ref{Eq:SSE-LB}), and upper bounds (U), according to
(\ref{Eq:SSE-UB}),~on the energy eigenvalue $E$ of the ground state
[$(n,\ell)=(1,0)$] of the spinless Salpeter equation with Coulomb-plus-linear
potential $V(r)=v(-a/r+br),$ for $a=0.2,$ $b=0.5,$ $m=\beta=1.$ The lower
bound is given by the general result (\ref{Eq:SSE-LB}) with $P(m=1)$ computed
from Eq.~(\ref{Eq:P(m)}), evaluated for mass $m=1$ and the funnel-potential
coupling strengths $c_1=va$ and $c_2=vb.$ For comparison, a (very accurate)
Rayleigh--Ritz variational upper bound $E$ is
depicted~too.}\label{Fig:SSE-bounds}\end{center}\end{figure}

Figure~\ref{Fig:SSE-bounds} shows, for a Coulomb-plus-linear potential of the
form $V(r)=v(-a/r+br),$~the lower and upper bounds on the lowest energy
eigenvalue $E$ of the spinless Salpeter equation, given by the envelopes of
the lower and upper families of tangential energy curves (\ref{Eq:SSE-LB})
and (\ref{Eq:SSE-UB}), as functions $E(v)$ of the ``overall'' coupling
parameter $v$ which multiplies the potential shape $-a/r+br.$ Again we
compare these bounds with the ground-state energy curve $E(v),$ obtained by
the Rayleigh--Ritz variational technique \cite{Reed78} with the Laguerre
basis states \cite{Lucha97}.

\section{Summary and conclusion}In this analysis we have studied the discrete
spectrum of semirelativistic ``spinless-Salpeter'' Hamiltonians $H,$ defined
in Eq.~(\ref{Eq:SH}), by an approach which is based principally on convexity.
We have at our disposal very definite information concerning, on the one
hand, the bottom of the spectrum of $H$ for the Coulomb potential,
$h(r)=-1/r,$ and, on the other hand, the entire spectrum of $H$ for the
harmonic-oscillator potential, $h(r)=r^2.$ The class of potentials that are
at the same time a convex transformation of $-1/r$ and a concave
transformation~of $r^2$ includes, for example, arbitrary linear combinations
of Coulomb, logarithmic, linear,~and harmonic-oscillator terms. In order to
obtain information about the eigenvalues $E$ of $H$ for arbitrary members
within this class of potentials, we have extended the---for nonrelativistic
Schr\"odinger operators well-established---formalism of envelope theory to
Hamiltonians with relativistic kinetic energies. The envelope technique
applied here takes advantage of the fact that all ``tangent lines'' to the
interaction potential $V(r)=g(h(r))$ in $H$ are potentials~of~the form
$ah(r)+b,$ and that, by convexity and the comparison theorem recalled in
Subsec.~\ref{Subsec:SCT}, the energy eigenvalues corresponding to these
``tangent'' potentials provide rigorous bounds to the unknown exact
eigenvalues $E$ of $H.$ If $e(v)$ denotes the energy function---or a suitable
bound to it---corresponding to the problem posed by a ``basis'' potential
$vh(r),$ where $v$~is~a positive coupling parameter, the envelopes of upper
and lower families of energy curves~may be found with the help of the
``principal envelope formula''$$E\approx\min_{v>0}[e(v)-ve'(v)+g(e'(v))]\
.$$Here, a sign of approximate equality is used to indicate that, for a
definite convexity~of~$g(h),$ the envelope theory yields lower bounds for
convex $g(h)$ and~upper bounds for concave $g(h).$ With the above principal
envelope formula at hand, all new spectral pairs $\{h(r),e(v)\}$ which may
become available at some future time can immediately be used to enrich our
collection of energy bounds. If the basis potential~$h(r)$ is a pure power,
these bounds can be written~as$$E_{n\ell}\approx\min_{r>0}
\left[\beta\sqrt{m^2+\frac{P_{n\ell}^2}{r^2}}+V(r)\right],$$where the numbers
$P_{n\ell}$ are obtained from the corresponding underlying basis
problems.~The power of this technique is illustrated, in Sec.~\ref{Sec:FP},
by our application to the ``funnel'' potential, $V(r)=-c_1/r+c_2r.$ For this
problem, we have employed both the semirelativistic Coulomb and
harmonic-oscillator problems to calculate, respectively, lower and upper
bounds on the energy eigenvalues of the spinless Salpeter equation.

We expect that such results would provide bounds on the energy eigenvalues
for general theoretical discussions, or be used as guides for more tightly
focussed analytic or numerical studies of the spectra of semirelativistic
``spinless-Salpeter'' Hamiltonians.

\section*{Acknowledgement}Partial financial support of this work under Grant
No.~GP3438 from the Natural Sciences and Engineering Research Council of
Canada, and the hospitality of the Erwin Schr\"odinger International
Institute for Mathematical Physics in Vienna is gratefully acknowledged by
one of us (R.~L.~H.).

\end{document}